\documentclass{PoS}

\usepackage{amsmath} 
\usepackage{amssymb}
\usepackage[utf8]{inputenc}

\title{New physics facing LFU and LFV tests in B physics}

\ShortTitle{NP facing LFU and LFV in $B$ physics}

\author{\speaker{Nejc Ko\v snik}\\
        Department of Physics, University of Ljubljana\\
        and\\
        Jo{\v z}ef Stefan Institute, Ljubljana\\
        E-mail: \email{nejc.kosnik@ijs.si}}

      \abstract{Lepton flavor universality in the Standard Model and
        hints of its violation in neutral ($R_K$) and charged currents
        $(R_{D^{(*)}})$ in $B$ decays are presented. We discuss model
        independent features as well as several leptoquark scenarios
        which are well suited to fit current values of $R_K$ and/or $R_{D^{(*)}}$.}

\FullConference{Fourth Annual Large Hadron Collider Physics\\
		13-18 June 2016\\
		Lund, Sweden}

\newcommand {\E}[1]{\times 10^{#1}}	
\newcommand {\e}[1]{\mathrm{~#1}}       
\newcommand{\mc}[1]{\mathcal{#1}}

\newcommand{\mrm}[1]{\mathrm{#1}}
\newcommand{\re}[0]{\mrm{Re}}
\newcommand{\im}[0]{\mrm{Im}}

\begin{document}

\section{Introduction}
Lepton flavor universality is present on the level of Standard
Model~(SM) gauge couplings and is broken by the Higgs Yukawa
couplings. Due to tiny neutrino masses and experimental
blindness to neutrino flavors the PMNS mixing matrix can be safely
neglected in $e^+ e^-$ or hadron collider. Lepton flavor universality (LFU) ratios,
constructed as ratios of process rates that differ only in charged
lepton flavor, are particularly clean quantities. This statement holds
both on the experimental front, where common systematics of the two
processes cancels out to some extent, as well as on the theoretical
side where
parametric uncertainties (e.g., CKM elements) and/or hadronic
parameter uncertainties also cancel. In some cases the SM prediction
of the LFU ratio is entirely determined by the charged lepton masses
taking part in the process. Consider, e.g., the leptonic decay width
induced by a $W$-exchange
\begin{equation}
  \label{eq:leptonic}
  \Gamma_{P(\bar q_j q_i) \to \ell \nu} = \frac{G_F^2 |V_{ij}|^2 f_P^2 m_P}{8\pi} m_\ell^2 \left(1- \frac{m_\ell^2}{m_P^2}\right)^2.
\end{equation}
When LFU ratio is taken with the above expression, the dependence on
the CKM element
vanishes, as well as on the hadronic decay constant (this latter feature
is specific to two-body decays)\footnote{The ratio receives
  small electromagnetic corrections~\cite{Cirigliano:2007xi}}.
LFU ratios have been repeatedly tested in meson decays and the results
in Tab.~\ref{tab:LFUtests} show remarkeble agreement with SM
expectations, where disparate charged lepton masses are the only source of LFU breaking.
\begin{table}[!htbp]
  \centering
  \begin{tabular}{|c||c|c|}\hline
    & SM & Experiment\\\hline
    $R^\pi_{e/\mu} = \Gamma_{\pi \to e \nu}/ \Gamma_{\pi \to \mu \nu}$& $1.2352(1) \E{-4}$& $1.2327(23)\E{-4}$\\\hline
    $R^K_{e/\mu}=\Gamma_{K \to e \nu}/ \Gamma_{K \to \mu \nu}$&
                                                                $2.477(1) \E{-5}$& $2.488(10)\E{-5}$\\\hline
    $R^K_{\tau/\mu}=\Gamma_{\tau \to K \nu}/ \Gamma_{K \to \mu \nu}$&
                                                                      $1.1162(3) \E{-2}$& $1.101(16)\E{-2}$\\\hline
    $R^B_{\tau/\mu}=\Gamma_{B \to \tau \nu}/ \Gamma_{B \to \mu \nu}$&
                                                                      $223$&
                                                                             $\gtrsim 100$\\\hline
  \end{tabular}
  \caption{Comparison of SM predictions and measured values of LFU
    ratios. Experimental values from~\cite{Agashe:2014kda}.}
  \label{tab:LFUtests}
\end{table}
However, charged current processes are not the most suitable for
catching potential beyond the large SM effects. Less SM
background is expected in neutral current processes which are
predicted to be strongly suppressed in the absence of exotic NP effects.

\section{Neutral current LFU: $R_K$}
The LHCb experiment reported an interesting result 
on the LFU $\mu/e$ ratio in $b \to s \ell^+ \ell^-$ process~\cite{Aaij:2014ora}:
\begin{equation}
\label{e1}
R_K = \frac{ \mc{B}( B \to K \mu^+ \mu^-)}{\mc{B}( B \to K e^+ e^-)}\Big|_{q^2 \in
    [1,6]\e{GeV^2}} = 0.745 \pm^{0.090}_{0.074} \pm 0.036  \quad \textrm{(LHCb)},
\end{equation}
about $2.6\sigma$ lower than the SM prediction
$R_K^{SM} =1.00(3)$\cite{Hiller:2003js,Bordone:2016gaq}. The
theoretical uncertainty is dominated by the electromagnetic radiative
corrections~\cite{Bordone:2016gaq}. Investigation of the
$B \to K^{(*)} \mu\mu$ process reveals deviations from SM prediction
in angular observable $P_5'$ in $B \to K^{*} \mu\mu$ as well as lower
than expected differential spectra in various decay modes
($B \to K^{*} \mu\mu$, $B \to K \mu\mu$,
$B \to \phi \mu \mu$)~\cite{Aaij:2014pli,Aaij:2015esa} which suggests
that the measured $R_K$ could be due to NP contributions in muonic
decay modes.

A sensible starting point in studies of low-energy phenomenology of
any NP model with degrees of freedom heavier than the weak scale is
the SM, complemented at the electroweak scale by
mass-dimension 6
operators (SM-EFT)~\cite{Buchmuller:1985jz,Grzadkowski:2010es,Alonso:2013hga}.
More convenient for the treatment of low energy neutral current processes is an effective
Hamiltonian that is matched onto to the SM-EFT through renormalization
group~(RG) running due to the full SM group above the electroweak
scale~\cite{Alonso:2013hga} and due to strong and electromagnetic RG
effects below the electroweak
scale~\cite{Buchalla:1995vs,Feruglio:2016gvd},
\begin{equation}
  \mc{L}_\mrm{eff} = \frac{4G_F}{\sqrt{2}} V_{tb} V_{ts}^* \left[ \sum_{i=1}^6 C_i \mc{O}_i
+\sum_{i=7,8,9,10,S,P} \left(C_i \mc{O}_i + C_i' \mc{O}'_i\right)\right].
\end{equation}
Among all the operators we will list only the semileptonic ones, which
carry a lepton index and may be responsible for lepton universality violation:
\begin{equation}
  \begin{split}
\mc{O}_9^{(\prime)} &= \frac{e^2}{(4\pi)^2} (\bar s \gamma_\mu P_{L(R)} b) (\bar \ell
\gamma^\mu \ell),\qquad \quad\! \mc{O}_{10}^{(\prime)} = \frac{e^2}{(4\pi)^2} (\bar s \gamma_\mu P_{L(R)} b) (\bar \ell
\gamma^\mu \gamma_5 \ell) ,\\
\mc{O}_S^{(\prime)} &= \frac{e^2}{(4\pi)^2} (\bar s P_{R(L)} b) (\bar \ell
\ell),\qquad \quad\!\qquad \, \mc{O}_{P}^{(\prime)} = \frac{e^2}{(4\pi)^2} (\bar s  P_{R(L)} b) (\bar \ell
\gamma_5 \ell). 
  \end{split}
\end{equation}
In the following we will only consider modification of the processes
with muons in the final state and accordingly we set $\ell = \mu$ in
the above operators, while the $b \to s ee$ interactions will be
assumed SM-like. The values of the SM Wilson coefficients at scale $m_B$
are $C_9^\mathrm{SM} = -C_{10}^\mathrm{SM} = 4.2$.  Furthermore, the SM-EFT matching onto the low energy Hamiltonian for
$b \to s \ell \ell$ processes already predicts that no tensor
operators would be generated, and that scalar operators are related to
pseudoscalars~\cite{Alonso:2014csa,Alonso:2015sja,Cata:2015lta,Feruglio:2016gvd},
$C_S = -C_P$, $C_S' = C_P'$. In the following $C_i$ will refer to the
value of the Wilson coefficient relative to its SM value.

The LFU universality $R_K$ itself does not allow explanation in terms
of scalar and pseudoscalar operators, $\mc{O}_S^{(')}$ and
$\mc{O}_P^{(')}$, since their size needed would cause excessive
$B_s \to \mu\mu$ branching fraction. One scenario that fits both $R_K$
and the rest of $b\to s \mu\mu$ data is the left-handed current
scenario with $C_9 = -C_{10}$~\cite{Hiller:2014yaa,Hiller:2014ula}, however this
is not the only possibility. Vector lepton currents $C_9$, $C_9'$, induced by $Z'$
coupled to muons and taus have been studied
in~\cite{Altmannshofer:2014cfa,Altmannshofer:2015mqa}.  Scenario with right-handed quark
current, $C_9' = -C_{10}'$, can address $R_K$, and can be potentially
resolved from the
left-handed scenario in other LFU
ratios~\cite{Hiller:2014ula,Altmannshofer:2013foa,Becirevic:2015asa}. 

\begin{figure}[!htbp]
  \centering
  \includegraphics[width=0.5\textwidth]{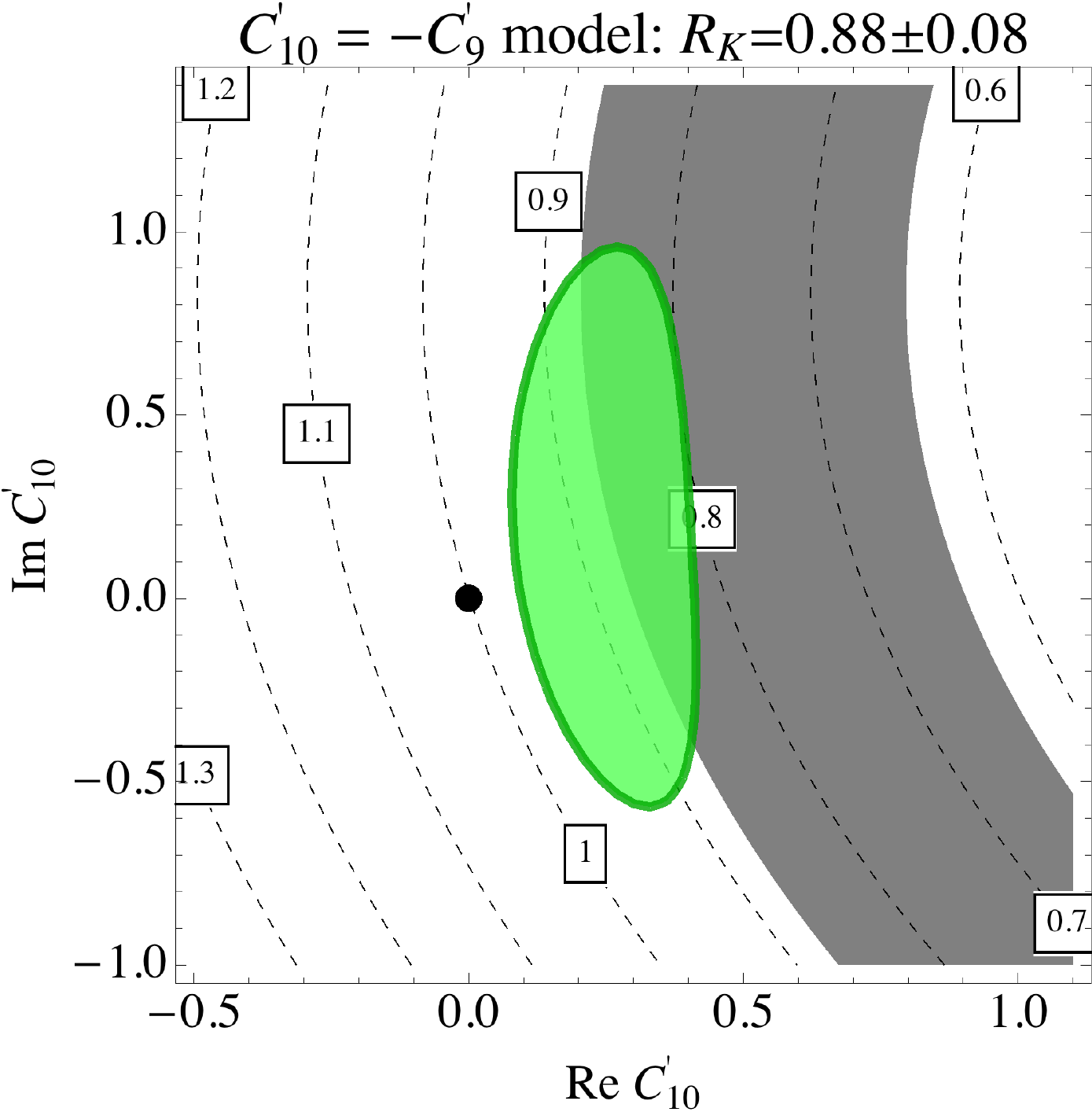}
  \caption{$R_K$ are denoted by dashed
    lines. Gray region represents the LHCb measurement of 
    $R_K$. Green
    contour is the $1\sigma$ fitted region to $\mc{B}(B_s \to \mu^+ \mu^-)$ and
    $\mc{B}(B \to K \mu^+ \mu^-)$.}
  \label{fig:RK}
\end{figure}

The analysis of
the $C_9' = -C_{10}'$ scenario was carried out
in~\cite{Becirevic:2015asa}. The experimental constraints that have been imposed
are the semileptonic and leptonic decay rates:
\begin{equation}
  \label{eq:3}
\begin{split}
   \mc{B}(B^+ \to K^+\mu^+ \mu^-)|_{q^2\in [15,22]\mrm{GeV}^2} &= (8.5
  \pm 0.3 \pm 0.4)\E{-8}\qquad \cite{Aaij:2014pli},\\
  \mc{B}(B_s \to \mu^+ \mu^-) &= (2.8_{-0.6}^{+0.7})\E{-9}
  \qquad \cite{CMS:2014xfa}.
\end{split}  
\end{equation}
The hadronic form factors that were employed had been calculated in unquenched
lattice QCD simulation~\cite{Bouchard:2013eph}, while for the $B_s$ decay
constant the FLAG average~\cite{Aoki:2013ldr} was used.
Prediction of $R_K$ cancels out form factor errors
and reads
\begin{equation}
\label{eq:rk-formula}
R_K (C_{10}^{\prime }) = 1.001(1) - 0.46\  \re [C_{10}^{\prime}] - 0.094(3)\  \im [C_{10}^{\prime}] +
0.057(1) |C_{10}^{\prime}|^2,
\end{equation}
where the remaining uncertainties are indicated by the numbers in parentheses.
In Fig.~\ref{fig:RK} we show contours of constant $R_K$ in the $C_{10}'$ plane 
using the formula~\eqref{eq:rk-formula}. Mapping the fitted region (green) to
$R_K$ we obtain the prediction
\begin{equation}
  \label{eq:RKpred}
  R_K^\mrm{pred.} = 0.88 \pm 0.08\,,
\end{equation}
which is indeed in good agreement with the LHCb measurement.
Possible UV completion of the $C_9' = -C_{10}'$ model is a
scalar leptoquark $\Delta$ in the SM representation $(3,2,1/6)$.
The defining Lagrangian contains a leptoquark (LQ) scalar field $\Delta$ that
is coupled to the SM fermions in Yukawa terms:
\begin{equation}
 \label{eq:Lag16}  
\begin{split}
  \mc{L} &= Y_{ij} \overline L_i \, i\tau^2 \Delta^* d_{Rj} 
+ \mrm{h.c.} \\
&= Y_{ij} \left(-\bar \ell_{Li} d_{Rj} \Delta^{(2/3)*} 
+ \bar \nu_{Lk}(V^\mrm{PMNS})^\dagger_{ki} d_{Rj} \Delta^{(-1/3)*} 
\right) + \mrm{h.c.}.
\end{split}
\end{equation}
At the scale $m_B$ the Wilson coefficients are expressed in terms of
LQ Yukawas
\begin{equation}
\label{C10LQ}
C_{10}^{\prime} = -C_{9}^{\prime}  = \frac{\pi}{2\sqrt{2} G_F
  V_{tb} V_{ts}^* \alpha}\, \frac{Y_{\mu b} Y_{\mu s}^*}{ m_\Delta^2}.
\end{equation}
Mass of $\Delta$ from direct searches at the LHC should be larger than
few 100's of GeV, with precise limit depending on its pattern of decay
branching fractions. For indirect effects that we are interested in
the precise mass is not an issue since it is the combination of Yukawa
couplings and masses that enters predictions here and all amplitudes
(with the exception of neutral meson mixing) scale as
$\sim Y^2/M_\Delta^2$. From different scaling in meson
$B_s$--$\bar B_s$ mixing we conclude that $\Delta$ should be lighter
than $\sim 100\e{TeV}$ in order to stay in the perturbative
regime. In~\cite{Becirevic:2015asa} we have predicted, alongside $R_K$, slightly
enhanced $B \to K^{(*)} \nu \nu$ decay mode and enhanced LFU ratio
$R_{K^{*}} = 1.11(8)$.

Further LQ scenarios have been proposed in the recent literature. A
weak triplet state $(\bar 3, 3,1/3)$ that implements $C_9 = -C_{10}$
effective theory has been presented in~\cite{Hiller:2014ula}. Weak singlet
state $(\bar 3, 1,1/3)$ was proposed in~\cite{Bauer:2015knc} to explain $R_K$ and alongside
with it also the charged LFU puzzle, $R_{D^{(*)}}$,
to be discussed in the next section. Both LQ states with hypercharge $1/3$ have
additional diquark couplings which must be suppressed to maintain
stability of the proton. Furthermore, viability of
$(\bar 3, 1,1/3)$ scenario due to additional flavor constraints has
been questioned in~\cite{Becirevic:2016oho}. A vector leptoquark
scenario in representation $(3,3,2/3)$ that addresses $R_K$ and
$R_{D^{(*)}}$ will be presented below.

\section{Charged current LFU: $R_{D^{(*)}}$}
In the charged-current induced semileptonic $B$ decays
we are witnessing persistent indications of disagreement with
the SM prediction of lepton flavor universality~(LFU) ratio in the
$\tau/\mu$ and/or $\tau/e$ sector. Namely, in the ratio
$R_{D^{(*)}} = \tfrac{\Gamma(B \to D^{(*)} \tau^- \bar\nu)}{\Gamma(B
  \to D^{(*)} \ell^- \bar\nu)}$ 
the
deviation from the SM is at $4\sigma$
level~\cite{Amhis:2014hma,Blake:2016olu}. The significance of the
discrepancy is driven by the experimental world
average~\cite{Amhis:2014hma},
\begin{equation}
  R_D^\mrm{exp} = 0.397\pm 0.040\pm 0.028,\qquad
R_{D^*}^\mrm{exp} = 0.316 \pm  0.016  \pm 0.010
\end{equation}
with a correlation coefficient of $-0.21$. The SM predictions are much
more precise:
\begin{equation}
R_D^\mrm{SM} = 0.297 \pm 0.017,\qquad R_{D^*}^\mrm{SM} = 0.252 \pm 0.003,
\end{equation}
and have been obtained in~\cite{Na:2015kha} and \cite{Fajfer:2012vx},
respectively. The $R_{D^{(*)}}$ puzzle has attracted a lot of
attention recently~(see e.g., \cite{Crivellin:2014zpa,Bhattacharya:2014wla,Bhattacharya:2015ida,Hati:2015awg,Sakaki:2014sea}).

In terms of the effective theory framework the vector ($g_V$), scalar
($g_S$), and tensor ($g_T$) operators modifying the semi-tauonic
decays give best fits to the measured $R_{D^{(*)}}$~\cite{Becirevic:2012jf,Freytsis:2015qca}:
\begin{equation}
\mc{L} = -\frac{4G_F V_{cb}}{\sqrt{2}} \Big[(1+g_V) (\bar \tau_L \gamma^\mu \nu_L)(\bar c_L \gamma_\mu b_L) 
+ g_S (\bar \tau_R \nu_L)(\bar c_R b_L) + g_T (\bar \tau_R \sigma^{\mu\nu}\nu_L)(\bar c_R \sigma_{\mu\nu}b_L)\Big] .
\end{equation}
Since the measured LFU violation requires large enhancement of the SM tree-level process
it is compulsory to invoke a NP model contributing at
tree-level. For NP studies involving charged colorless scalars
see, e.g.,~\cite{Crivellin:2012ye,Celis:2012dk,Ko:2012sv,Crivellin:2015hha}, whereas
colored states (leptoquarks) have been pursued in,
e.g.,~\cite{Dorsner:2013tla,Sakaki:2013bfa,Bauer:2015knc,Li:2016vvp}.

\begin{figure}[!htbp]
  \centering
  \begin{tabular}{cc}
 \includegraphics[width=0.39\textwidth]{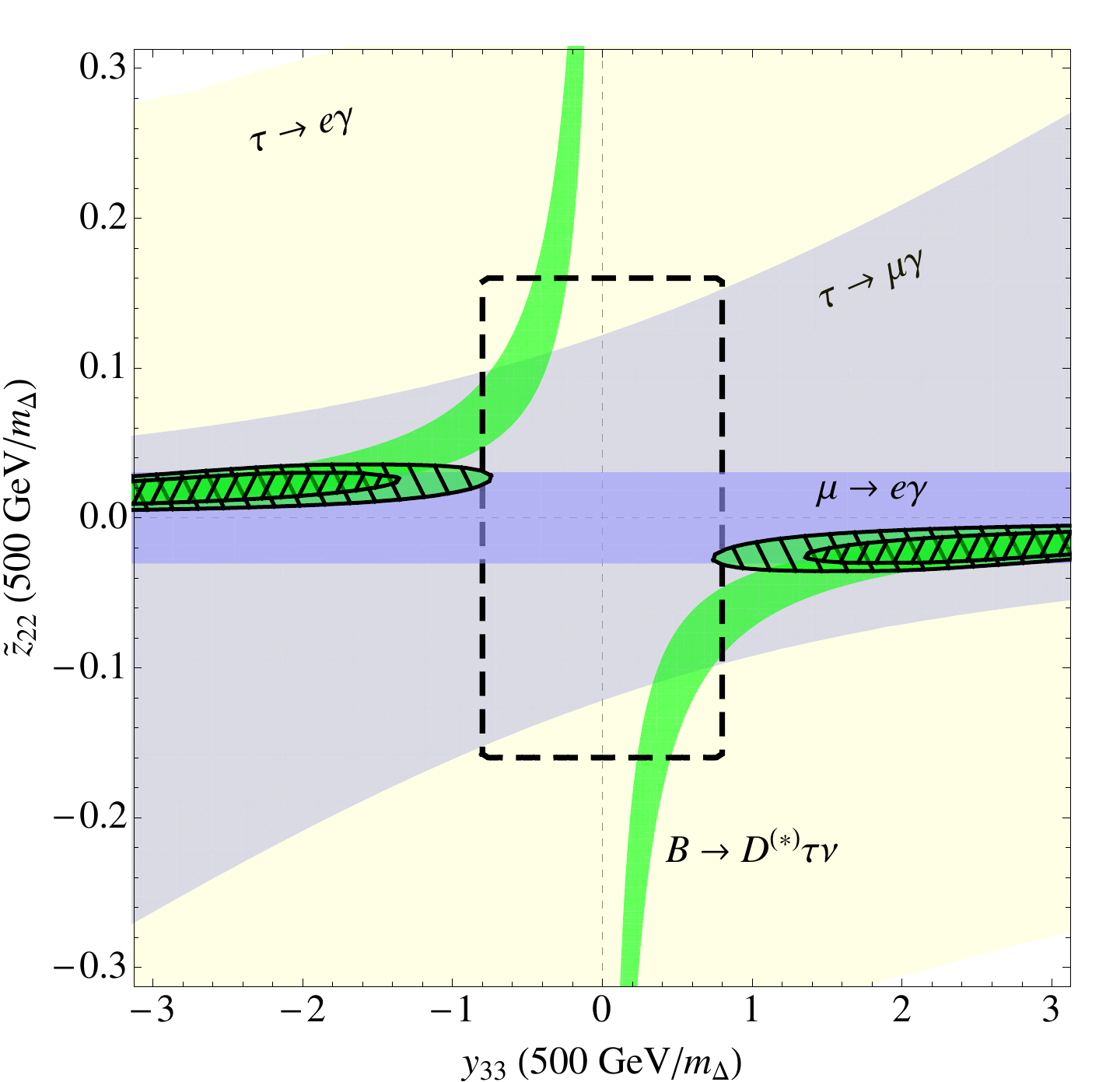}   & \includegraphics[width=0.62\textwidth]{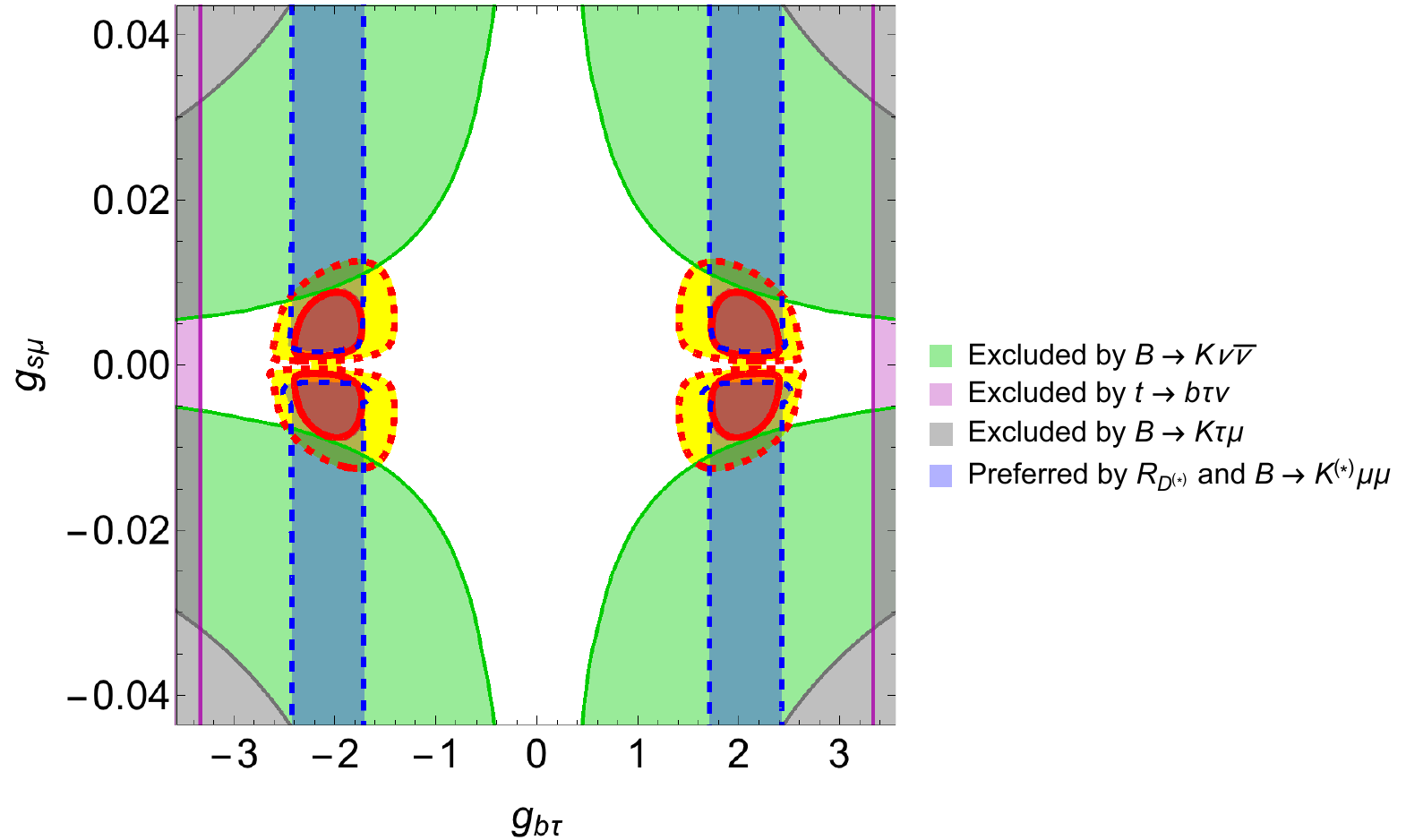}
  \end{tabular}
  \caption{Left panel: Constraints on the $R_2$ leptoquark scenario
    embedded in $SU(5)$ GUT theory~\cite{Dorsner:2013tla}. Dashed frame region ensures perturbativity of the Yukawa
  couplings all the way to the GUT scale. Hatched green area denotes the
  best fit region. Right panel: Constraints on the vector LQ $U_3$, as
discussed in~\cite{Fajfer:2015ycq}. Red and yellow regions are the
$1$- and $2$-$\sigma$ fitted regions.} 
  \label{fig:R2cons}
\end{figure}

First, consider a model with a scalar leptoquark $R_2$ in the representation
$(3,2,7/6)$ that couples non-chirally
\begin{equation}
  \mc{L}_\mrm{LQ} = \bar \ell_R Y R_2^\dagger Q + \bar u_R Z i\tau_2 R_2^T L  +\mrm{h.c.}.
\end{equation}
Here we have chosen $Y$ and $Z$ Yukawa matrices in a way to have
minimal number of non-zero elements and still allow explaining
$R_{D^{(*)}}$. $Y$ only couples $\tau_R$ to $b_L$, $Z$ couples $c_R$
to all neutrinos and charged leptons. It has been shown
in a realistic $SU(5)$ unification model~\cite{Dorsner:2013tla} that $\mu \to e \gamma$ is the most relevant
constraint that forces $Y_{\tau b}$ to be large and
consequently $R_2$ is well suited to direct searches in $\tau b$ final
states. Left panel in Fig.~\ref{fig:R2cons} outlines the effect of
various constraints.

The second scenario is the one with the vector LQ, $U_3
(3,3,2/3)$.
Such weak triplet always contributes to left-handed current operators,
both in charged and neutral currents, and in can turn interfere with
the SM contributions very effectively to address $R_{D^{(*)}}$ as well
as $R_K$~\cite{Fajfer:2015ycq}:
\begin{equation}
  \label{eq:LQLagComp}
  \begin{split}
      \mc{L}_{U_3} =\,\, &U^{(2/3)}_{3\mu} \, \Big[ (Vg
       )_{ij}\, \bar u_i \gamma^\mu P_L \nu_j  -
        g_{ij} \,\bar d_i \gamma^\mu P_L \ell_j \Big]\\
 &+U^{(5/3)}_{3\mu}\, (\sqrt{2}  V g)_{ij}\, \bar u_i \gamma^\mu P_L
 \ell_j \\
  &+U^{(-1/3)}_{3\mu}\, (\sqrt{2} g )_{ij}\, \bar d_i \gamma^\mu P_L
 \nu_j+\mrm{h.c.}.
  \end{split}
\end{equation}
Couplings are judiciously chosen in the down quark-charged lepton
sector:
\begin{equation}
\label{eq:texture}
g =
\begin{pmatrix}
  0 & 0 & 0\\
  0 & g_{s \mu} & 0\\
  0 & g_{b \mu} & g_{b \tau}
\end{pmatrix},
\qquad
V g =
\begin{pmatrix}
  0 & V_{us} g_{s\mu} + V_{ub} g_{b\mu} & V_{ub} g_{b\tau}\\
  0 & V_{cs} g_{s\mu} + V_{cb} g_{b\mu} & V_{cb} g_{b\tau}\\
  0 & V_{ts} g_{s\mu} + V_{tb} g_{b\mu} & V_{tb} g_{b\tau}
\end{pmatrix},
\end{equation} 
and are related to up-quarks and neutrino interactions by CKM rotation.

Again, $R_{D^{(*)}}$ requires large
coupling to the third generation matter. For this scenario it is the
$B \to K \nu \nu$ constraint that most effectively probes the LFV
couplings that also drive the $B \to K \tau \mu$, as seen in
right-panel in Fig.~\ref{fig:R2cons}.

\section{Conclusion}
The two recent LFU violating observables can be explained by
tree-level LQ contributions. We have demonstrated that a scalar
leptoquark $\Delta (3,2,1/6)$ can successfully explain $R_K$ whereas
$R_2 (3,2,7/6)$ can shift $R_{D^{(*)}}$ closer to
experiment. Finally, triplet vector leptoquark $U_3 (3,3,2/3)$ can
modify the left-handed current operators and explain both LFU puzzles.
In all the above cases we observe that explaining $R_{D^{(*)}}$ calls
for large coupling between $\tau$ and $b$. Lepton flavor violation
constraints generally do not allow for additional sizable LQ couplings
to fermions. In all the considered scenarios bounds on the LFV processes
provide the most stringent constraint, e.g. $\mu \to e \gamma$ for
$R_2 (3,2,7/6)$ and $B\to K \nu \nu$ for $U_3 (3,3,2/3)$.

\end{document}